\documentclass[11pt]{article}
\usepackage{a4wide}
\usepackage{epsfig,epsf,graphics,psfrag}
\usepackage{times}
\usepackage{float}
\usepackage{color}
\usepackage{amsfonts,amssymb,stmaryrd,latexsym,amsmath}
\usepackage{multirow}
\usepackage{array} 

\begin{document}
\title{Exploring the threshold behavior and implications on the nature of $Y(4260)$ and $Z_c(3900)$}

\author{Xiao-Hai Liu$^1$\footnote{{\it E-mail address:}liuxiaohai@pku.edu.cn},  \
Gang Li$^2$\footnote{{\it E-mail address:}gli@mail.qfnu.edu.cn}
     \\
     {\it\small$\rm ^1$ Department of Physics and State Key Laboratory of Nuclear Physics and Technology,}\\
     {\it\small Peking University, Beijing 100871, China}
     \\
    {\it\small$\rm ^2$Department of Physics, Qufu Normal University, Qufu
273165, P.R. China}
          }

\maketitle

\begin{abstract}
By assuming that $Y(4260)$ either is a $D_1\bar{D}$ molecular
state, or has sizeable couplings with $D_0 \bar{D}^*$ and
$D_1^\prime \bar{D}$, we investigate several decay modes of
$Y(4260)$. Under the special kinematic configurations, triangle
singularity (TS) may occur in the rescattering amplitude, which
will significantly change the threshold behavior. Obvious
threshold enhancements or narrow cusp structures appear quite
naturally without introducing a genuine resonance. We emphasize
that the radiative decay modes may be useful for studying
$D^{(*)}\bar{D}$ $S$-wave scattering. \noindent
\end{abstract}



\thispagestyle{empty}
\newpage


\section{Introduction}
The observation of $Y(4260)$ was first announced by BABAR
collaboration in 2005 \cite{Aubert:2005rm}. It was then confirmed
by CLEO \cite{He:2006kg} and BELLE \cite{Yuan:2007sj}. Because of
the puzzling characters of this charmonium-like particle, it has
attracted much attention both in experimental and theoretical
research. There is no direct correspondence of $Y(4260)$ in naive
quark model classifications. Furthermore, as a charmonium
candidate, its preferable decay mode would be open charm decays.
However, apart from the hidden charm decay channel, it has not
been observed in the $\bar{D}D$, $\bar{D}D^{*}$, $\bar{D}^*D^{*}$,
and $DD^*\pi$
modes~\cite{Pakhlova:2008zza,Aubert:2009aq,Pakhlova:2009jv,Abe:2006fj,Aubert:2006mi}.
The $R$-value scan at BES around 4.26 GeV appears to have a dip
instead of a bump structure~\cite{Ablikim:2007gd}. These peculiar
characters initiate lots of discussions about the nature of
$Y(4260)$, such as the explanations of hybrid state
\cite{Zhu:2005hp,Close:2005iz,Kou:2005gt}, tetraquark state
\cite{Ebert:2005nc,Ebert:2008kb,Maiani:2005pe},
$\Lambda_c\bar{\Lambda}_c$ baryonium state \cite{Qiao:2005av},
$\chi_{c0}\rho$ or $\chi_{c1}\omega$ molecular state
\cite{Liu:2005ay,Yuan:2005dr}, conventional $c\bar{c}$ state renormalized by $\chi_{c0}\omega$~\cite{Dai:2012pb},  $D_1\bar{D}$ or $D_0\bar{D}^*$
molecular state
\cite{Kalashnikova:2008qr,Close:2009ag,Close:2010wq,Ding:2008gr,Li:2013bca},
and so on.

The recent observation of BESIII revives the discussion on the
nature of $Y(4260)$. A charged charmonium-like resonance
structure, which is temporally named as $Z_c(3900)$, is observed
in the invariant mass spectrum of $J/\psi\pi^\pm$ from $Y(4260)\to
J/\psi\pi^+\pi^-$ \cite{Ablikim:2013mio}. If the observed
structure is a genuine particle, it obviously cannot be a
conventional $c\bar{c}$ state. Now there exist two unusual states
in this decay channel. This observation was confirmed by BELLE
\cite{Liu:2013dau}. A similar $Z_c$ was also confirmed by CLEO in
$\psi(4170)\to J/\psi\pi^+\pi^-$ \cite{Xiao:2013iha}. Together
with the observations of $Z(4430)$ in $\psi^\prime\pi^\pm$
\cite{Choi:2007wga}, $Z_{1,2}$ in $\chi_{c1}\pi^\pm$
\cite{Mizuk:2008me}, and $Z_b$ in $\Upsilon(nS)\pi^\pm$
($h_b(mP)\pi^\pm$)\cite{Belle:2011aa}, these unusual charged
states not only enrich the knowledge on hadron spectroscopy
largely, but also bring new challenges. Some recent discussions
about $Z_c(3900)$ can be found in
Refs.~\cite{Wang:2013cya,Guo:2013sya,Voloshin:2013dpa,Chen:2013coa,Liu:2013rxa,Wang:2013hga,Dong:2013iqa,Braaten:2013boa,Zhang:2013aoa}.

One remarkable character of these unconventional charmonium-like
resonance structures is that many of them are observed at the
thresholds of some charmed anti-charmed meson pairs. Therefore, to
some extent, it is reasonable to interpret them as molecular
states. There is another similar description. The couple channel
effect will play a role when investigating the pertinent hadron
spectrum. Especially, the couple channel effect might largely
affect the threshold phenomena, which has ever been taken as a
dynamical mechanism in explaining the observation of the charged
botomonium-like (charmonium-like)
structures~\cite{Chen:2011pv,Chen:2011zv,Bugg:2011jr,Chen:2011xk,Wang:2013cya,Chen:2013coa}.
In this work, we will assume that $Y(4260)$ either is a $D_1
\bar{D}$ molecular state, or has sizeable couplings with
$D_0\bar{D}^*$ and $D_1^\prime \bar{D}$. Here $D_1$ and $D_1^\prime$ refer to the narrow and
broad axial vector charmed mesons, i.e. $D_1(2420)$ ($\Gamma \simeq 27 \mbox{MeV}$)
and $D_1(2430)$ ($\Gamma \simeq 384 \mbox{MeV}$), respectively~\cite{Beringer:1900zz}. Under such assumptions, a
consistent description of many of the experimental observations
can be obtained, such as its non-obersvation in open charm decays,
or the observation of $Z_c(3900)$ as mentioned in
Ref.~\cite{Wang:2013cya}. We will mainly concentrate on the
discussion of the threshold phenomena that may result from these
assumptions in its strong and radiative decay channels.

\section{Threshold enhancement phenomena}
\begin{figure}[tb]
  \centering
  \includegraphics[width=0.8\hsize]{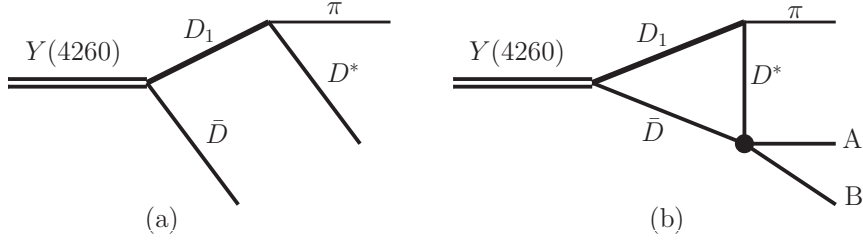}\\
  \caption{(a): Diagram of $Y(4260)$ main decay mode by supposing it is a $D_1\bar{D}$ molecular state. (b): the rescattering process, where A and B are some specified final states. The charge conjugate diagrams are implicit.}\label{rescattering}
\end{figure}

If $Y(4260)$ is a $D_1\bar{D}+c.c.$ molecular state, its main
decay channel would be $\bar{D}D^*\pi+c.c.$ (the charge conjugate
part $c.c.$ will be implicit sometimes for brevity), as
illustrated in Fig.~\ref{rescattering}(a). Since the momenta of
the produced $\bar{D}D^*$ will be very small, the final state
interactions (FSI) can be expected to play a role when analyzing
some decay processes. We illustrate the rescattering processes in
Fig.~\ref{rescattering}(b), where $A$ and $B$ are some specified
final particles, and the black bubble indicates some unknown
couplings. If FSI are strong, they may significantly change the
decay properties. Some
non-analytical structures of the transition amplitude such as the
cusp effect may occur. We will explore several pertinent decay
modes and discuss some interesting threshold behaviors in the
following sections.

\subsection{The model}

Throughout this paper we will adopt the heavy quark symmetry and
chiral symmetry. Within heavy hadron chiral perturbation theory
(HHChPT) \cite{Wise:1992hn,Casalbuoni:1996pg,Colangelo:2012xi},
the heavy meson multiplets are defined in the following way:
\begin{eqnarray}
 H_a  &=& \frac{1+{\rlap{v}/}}{2}[\mathcal{D}_{a\mu}^*\gamma^\mu-\mathcal{D}_a\gamma_5] \ , \\
  S_a &=& \frac{1+{\rlap{v}/}}{2} \left[\mathcal{D}_{1a}^{\prime \mu}\gamma_\mu\gamma_5-\mathcal{D}_{0a}^*\right] \ ,\\
   T_a^\mu &=&\frac{1+{\rlap{v}/}}{2} \left\{ \mathcal{D}^{\mu\nu}_{2a} \gamma_\nu -\sqrt{3 \over 2}\mathcal{D}_{1a\nu}  \gamma_5 \left[
g^{\mu \nu}-{1 \over 3} \gamma^\nu (\gamma^\mu-v^\mu) \right]
\right\} \ ,
\end{eqnarray}
where $H_a$ and $S_a$ ($T_a^\mu$) describe the $S$-wave and
$P$-wave heavy-light systems respectively, and $a$ is light flavor
index. Some pertinent effective Lagrangian describing the
interactions among these multiplets and Goldstone bosons according
to HHChPT read
\begin{eqnarray}
  {\mathcal L}_S &=&   ih\ Tr [{\bar H}_a S_b \gamma_\mu \gamma_5 {\cal
A}_{ba}^\mu ]\, + \ h.c. \ ,  \\
{\mathcal L}_T &=&  i{h^\prime \over \Lambda_\chi}\ Tr[{\bar H}_a
T^\mu_b \gamma^\lambda\gamma_5 ( D_\mu \mathcal{A}_\lambda +
D_\lambda \mathcal{A}_\mu)_{ba} ] + h.c.,   \label{efflag}
\end{eqnarray}
where
\begin{eqnarray}
  \mathcal{V}_\mu &=& \frac{1}{2} \left( \xi^\dag\partial_\mu\xi  + \xi\partial_\mu\xi^\dag  \right), \\
  \mathcal{A}_\mu &=& \frac{1}{2} \left( \xi^\dag\partial_\mu\xi  - \xi\partial_\mu\xi^\dag  \right), \\
  D_\mu &=& \partial_\mu +\mathcal{V}_\mu \ , \ \xi = e^{i\mathcal{M}/f_\pi}\ ,
\end{eqnarray}
and $\mathcal{M}$ is a $3\times 3$ hermitian matrix indicating the
octet of Goldstone bosons:
\begin{eqnarray}
\mathcal{M}= \left(
  \begin{array}{ccc}
    \frac{1}{\sqrt{2}}\pi^0+ \frac{1}{\sqrt{6}}\eta & \pi^+ & K^{+} \\
    \pi^- & -\frac{1}{\sqrt{2}}\pi^0+ \frac{1}{\sqrt{6}}\eta  & K^{0} \\
    K^{-} & \bar{K}^{0} & -\sqrt{\frac{2}{3}}\eta \\
  \end{array}
\right).
\end{eqnarray}
By assuming $Y(4260)$ is a $D_1\bar{D}$ molecule,  the Lagrangian
is constructed as
\begin{eqnarray}
\mathcal{L}_Y = g_Y Y_\mu (D_1^\mu \bar{D} - \bar{D}_1^\mu D ),
\end{eqnarray}
which is an $S$-wave coupling.

Concerning the rescattering process illustrated in
Fig.~\ref{rescattering}(b), we will discuss an interesting
singularity that may appear in the transition amplitude. I.e.,
with some special kinematic configurations, all of the three
intermediate states contained in the loop can be on-shell
simultaneously. This is the so called triangle singularity (TS) or
"two cut" condition \cite{Wu:2011yx,Wang:2013cya,Wang:2013hga}.
Since this kind of singularities usually appear when the mass of
the external particle is very close to the threshold of
intermediate states, it may change the threshold behavior
dramatically and show up directly as a bump in the
amplitude~\cite{Landshoff:1962,Eden:1966}.

The explicit values of the coupling constants are not well
determined for the moment. In order to obtain some less
model-dependent results, we will mainly pay attention to the
lineshape behavior of some pertinent invariant mass spectrum only.

\subsection{$Y(4260) \to \bar{D}D^*\pi$}

Taking $AB$ in Fig.~\ref{rescattering}(b) as $\bar{D}D^*$, we will
firstly discuss the $\bar{D}D^*\pi$ final states, and mainly
concentrate on the lineshape behavior of $\bar{D}D^*$ invariant
mass spectrum.

Concerning the $S$-wave rescattering process $\bar{D}D^*\to
\bar{D}D^*$, according to the power counting schemes discussed in
the Refs.~\cite{AlFiky:2005jd,Valderrama:2012jv}, the contact
interactions will stay at the leading order while one pion
exchange can be considered as a subleading order correction. We
will then just take into account the contact interaction for the
moment. The effective Lagrangian reads
\begin{equation}\label{4hcontact}
  \mathcal{L}_{4H}=C_1\ Tr[\bar{H}^Q H^Q \gamma_\mu]\ Tr[H^{\bar{Q}} \bar{H}^{\bar{Q}}  \gamma^\mu] +
  C_2\ Tr[\bar{H}^Q H^Q \gamma_\mu\gamma_5]\ Tr[H^{\bar{Q}} \bar{H}^{\bar{Q}}  \gamma^\mu \gamma_5].
\end{equation}
Taking into account the one pion exchange interaction will change
the triangle diagram of Fig.~\ref{rescattering}(b) into a box
diagram. However, it will not significantly change the character
of TS, because the singular properties of the box diagram could be
attributed to those of the triangle diagram ~\cite{Eden:1966}. As
discussed in Refs.\cite{Wang:2013cya,Wang:2013hga}, the triangle
diagram with a local contact interaction is consistent with the
non-local box diagram when the $\bar{D}D^*$ momenta are small.  As
a result, there will be not much discrepancy by analyzing the
triangle or box diagram when discussing the threshold phenomena
mentioned in this work. By utilizing the effective contact
interaction, we concentrate on the similar enhancement structure
around $3.9$ GeV resulting from the cusp effect instead of a
genuine state.

\begin{figure}[tb]
\centering
\begin{minipage}{8cm}\includegraphics[width=1.0\hsize]{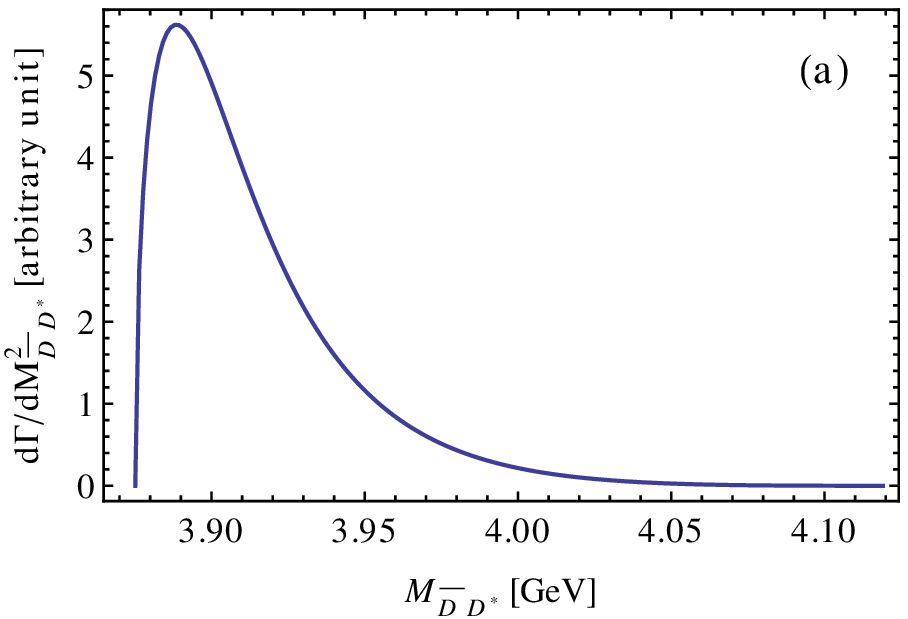}\end{minipage}\begin{minipage}{8cm}\includegraphics[width=1.0\hsize]{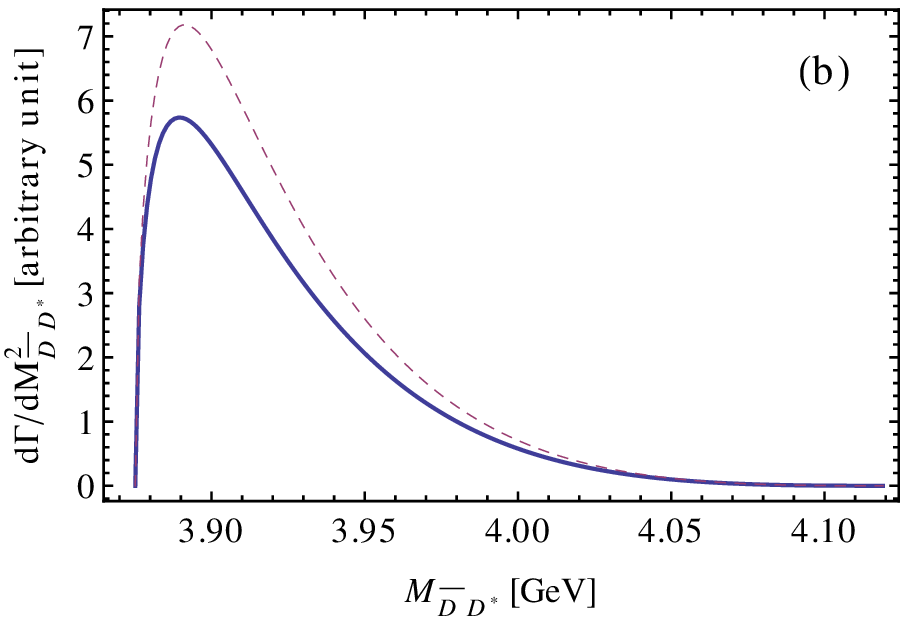}\end{minipage}
\caption{Invariant mass spectrum of $\bar{D}D^*$ in $Y(4260) \to
D_1 \bar{D} \to \bar{D} D^*\pi$. (a) and (b) correspond to tree
and loop diagrams respectively. And in (b), solid (dashed) line
corresponds to the result with (without) taking into account the
width of $D_1$.}\label{D1DDstr}
\end{figure}

Since $D_1\to D^*\pi$ is a D-wave decay while the $S$-wave decay
is forbidden according to heavy quark symmetry, which is one of
the main reason why this $D_1$ is so narrow. For the process
$Y(4260)\to \bar{D}D_1 \to \bar{D}D^*\pi$, the angular momentum
between the pion and $\bar{D}D^*$ system is D-wave. The amplitude
will be proportional to $|\bf{p}_\pi|^2$, where $\bf{p}_\pi$ is
the three momentum of pion meson. The partial width
$d\Gamma/dM_{\bar{D}D^*}$ will be proportional to (and very
sensitive to) $|\bf{p}_\pi|^5$. The threshold of the $\bar{D}D^*$
system exactly corresponds to $|\bf{p}_\pi|_{max}$. Although only
the tree diagram is considered, there appears an obvious threshold
enhancement on the $\bar{D}D^*$ distribution, as illustrated in
Fig.~\ref{D1DDstr}(a).

Since $Y(4260)$ just stay in the vicinity of the $D_1\bar{D}$
threshold, the momenta of $\bar{D}$ and $D^*$ via $Y(4260)\to
\bar{D} D_1$ decays are small. And from the above discussion it
could be concluded that most of the $\bar{D}D^*$ events will be
accumulated in the near threshold region, where we may expect
strong FSI. According to the above effective Lagrangian, the
transition amplitude corresponding to the rescattering process
Fig.~\ref{rescattering}(b) reads
\begin{eqnarray}\label{loopampc0}
  T_{Y\to\pi D^* \bar{D}}^{\mbox{loop}} &=& ig_c \int \frac{d^4 l}{(2\pi)^4} \frac{3p_\pi \cdot \epsilon_Y p_\pi \cdot \epsilon_{D^*} + ((v\cdot p_\pi)^2 - p_\pi^2)
  \epsilon_Y \cdot \epsilon_{D^*} }{(l^2 -m_{D}^2)((P-l)^2 -m_{D_1}^2)((P-p_\pi-l)^2 -m_{D^*}^2)},
\end{eqnarray}
where $g_c$ is the combination of some relevant coupling
constants, $\epsilon_Y$ and $\epsilon_{D^*}$ are the polarization
vectors of $Y(4260)$ and $D^*$ respectively, and $P$ is the
momentum of the initial particle. The sums of the polarizations for the heavy vector mesons
are $\sum \epsilon^\mu \epsilon^{*\nu}=-g^{\mu\nu}+v^\mu v^\nu$. In Eq.~(\ref{loopampc0}), there
is no integral momentum appeared in the nominator. It
is therefore just a scalar three point function, which can be
taken as a form factor for the $Y\psi\pi\pi$ coupling. The two-cut
condition indicates the singularity arising from the triangle
diagram in the form factor coupling. The numerical result
corresponding to the rescattering diagram is displayed in
Fig.~\ref{D1DDstr}(b), where an obvious threshold enhancement is
also observed. However, we have not taken into account the
interference between the tree and loop diagram, since there are
some unknown coupling constants. It is also not easy to judge
quantitatively the relative strength between the loop and tree
diagram for the moment. It should be mentioned some other
rescattering amplitudes share the similar formula with
Eq.~(\ref{loopampc0}). We will omit them in the following
discussions for brevity.

The doublet  $D_0$ and $D_1^\prime$ combined in $S_a$ of Eq.~(2)
are too broad to form a relatively narrower molecular state.
However, since the thresholds of $D_0 \bar{D}^*$ and $D_1^\prime
\bar{D}$ are very close to the mass of $Y(4260)$, it is still
justifiable to assume larger couplings of $Y(4260)$ with these
combinations. We will try to explore the lineshape behaviors under
such assumptions in the following. The threshold of $D_2\bar{D}$
is also close to $Y(4260)$. But their coupling is of $D$-wave,
which should be suppressed compared with other combinations. We
will not discuss it in this work.

\begin{figure}[tb]
\centering
\begin{minipage}{8cm}\includegraphics[width=1.0\hsize]{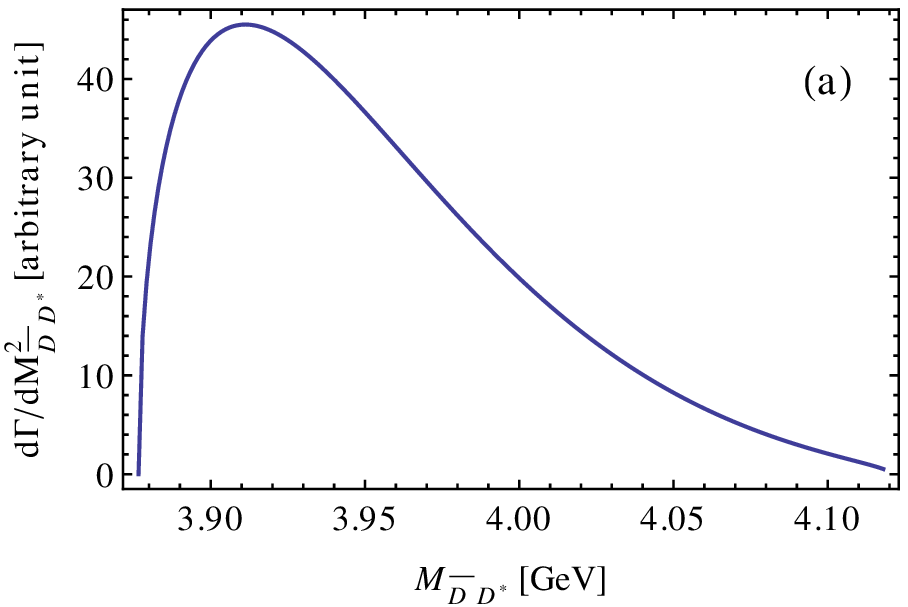}\end{minipage}\begin{minipage}{8cm}\includegraphics[width=1.0\hsize]{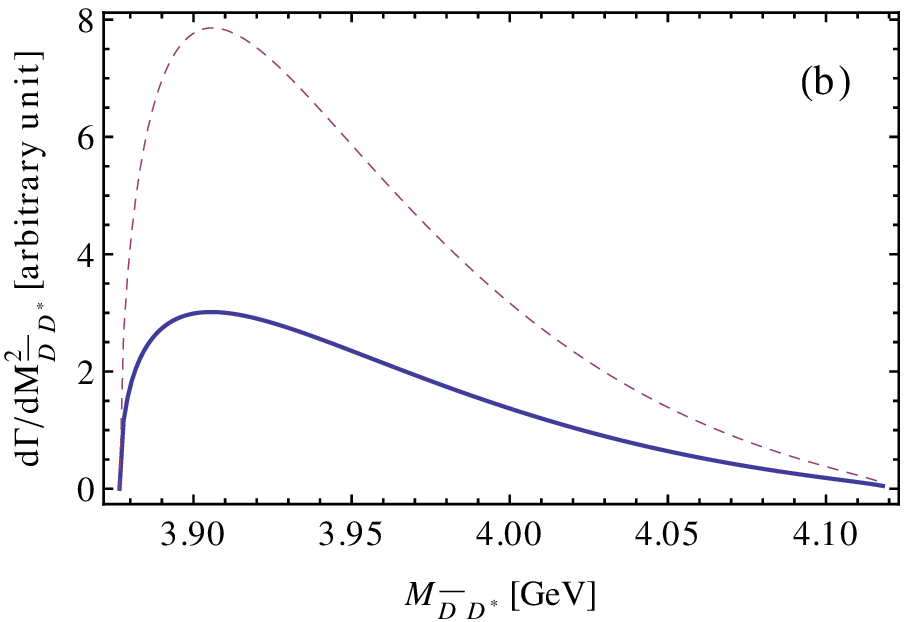}\end{minipage}
\caption{Invariant mass spectrum of $\bar{D}D^*$ in $Y(4260) \to
\bar{D}_0 D^* \to \bar{D} D^*\pi$. (a) and (b) correspond to tree
and loop diagrams respectively. And in (b), solid (dashed) line
corresponds to the result with (without) taking into account the
width of $D_0$.}\label{D0DDstar}
\end{figure}

\begin{figure}[tb]
\centering
\begin{minipage}{8cm}\includegraphics[width=1.0\hsize]{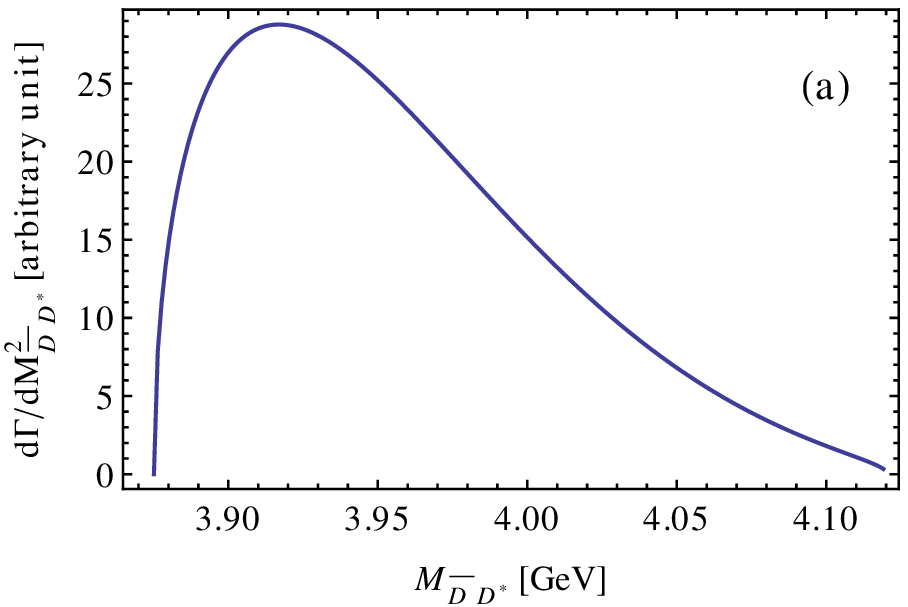}\end{minipage}\begin{minipage}{8cm}\includegraphics[width=1.0\hsize]{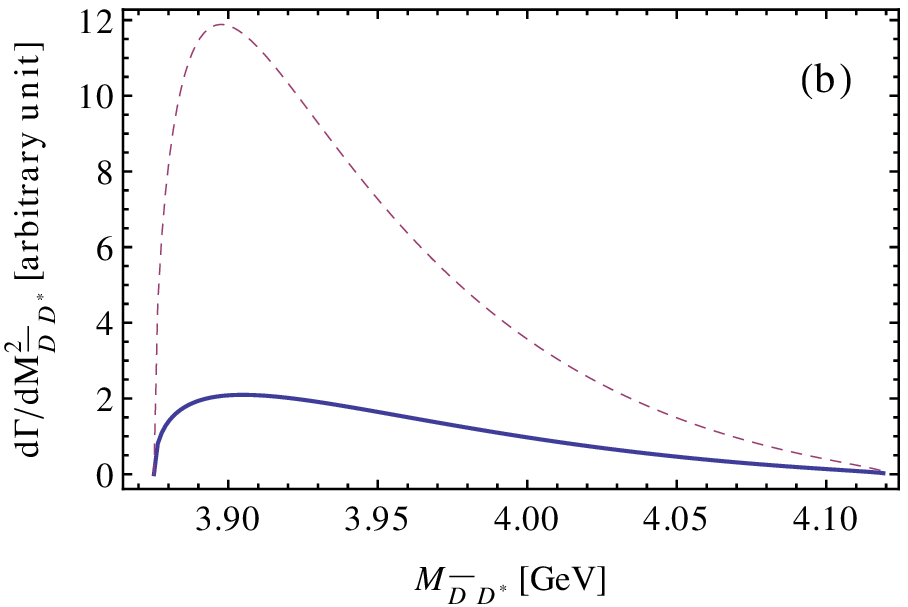}\end{minipage}
\caption{Invariant mass spectrum of $\bar{D}D^*$ in $Y(4260) \to
D_1^\prime \bar{D} \to \bar{D} D^*\pi$. (a) and (b) correspond to
tree and loop diagrams respectively. And in (b), solid (dashed)
line corresponds to the result with (without) taking into account
the width of $D_1^\prime$.}\label{D1primeDDstar}
\end{figure}

$D_0 \to D\pi$ and $D_1^\prime\to D^*\pi$ are $S$-wave decays.
Therefore only the tree diagram itself can not lead to obvious
threshold enhancement structures, as illustrated in
Fig.~\ref{D0DDstar}(a) and Fig.~\ref{D1primeDDstar}(a). When
evaluating the loop diagram, since the intermediate states are
very broad, we also need to take into account the influence of the
larger width. For estimating this, the propagator $l^2-m^2$ will
be changed into $l^2-m^2+im\Gamma$ in Eq.~\ref{loopampc0}, where
$\Gamma$ is the decay width of the corresponding state with mass
$m$. However, it should be mentioned this tentative prescription
of considering width effect is not quite justifiable since it will
destroy unitarity. As a qualitative analysis, we will ignore this
defect for the moment. From Fig.~\ref{D0DDstar}(b) and
Fig.~\ref{D1primeDDstar}(b), we can see that it will lower the
rescattering amplitude when considering the broad width influence.
And even taking into account rescattering processes, it seems that
the threshold enhancement behavior is still not significant
compared with the suitation in  $Y(4260) \to D_1 \bar{D} \to
\bar{D} D^*\pi$ process. But they may offer some background when
analyzing these decay channels.

\section{ $Y(4260)\to J/\psi(\psi^\prime) \pi\pi$}

\begin{figure}
  \centering
  \includegraphics[width=0.8\hsize]{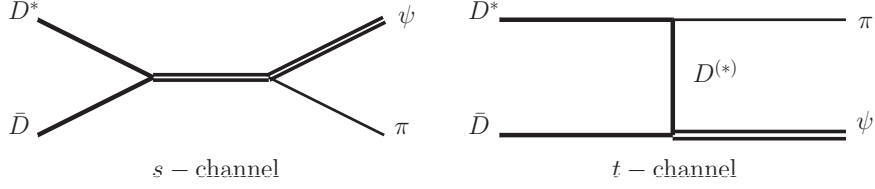}\\
  \caption{The rescattering process $\bar{D}D^*\to \psi\pi$ may receive contributions from $s$-channel and $t$-channel.}\label{stchannel}
\end{figure}

Taking $AB$ appeared in Fig.~\ref{rescattering}(b) as
$J/\psi(\psi^\prime)\pi$, we will then discuss the
$J/\psi(\psi^\prime) \pi\pi$ final states, and some of the points
have also been mentioned in Ref.~\cite{Wang:2013cya}. Concerning
the rescattering process $\bar{D}D^*\to J/\psi(\psi^\prime)\pi$,
generally speaking, it will receive contributions from $s$-channel
and $t$-channel as depicted in Fig.~\ref{stchannel}. However as we
argued in the previous section, the $t$-channel will correspond to
the box diagram, which will not significantly change the threshold
behavior compared with the triangle diagram. And we hope to
discuss the cusp structure without introducing a genuine resonance
close to the $\bar{D}D^*$ threshold apparently. For qualitative
estimation an effective contact interaction is constructed as
\begin{eqnarray}
\mathcal{L}_\psi =g_\psi \psi^\mu(D^*_\mu\bar{D}-\bar{D}^*_\mu
D)\pi,
\end{eqnarray}
where we assume an $S$-wave coupling, and it means the quantum
number of $\psi\pi$ system will be $J^P=1^+$.

\begin{figure}[tb]
\centering
\begin{minipage}{8cm}\includegraphics[width=1.0\hsize]{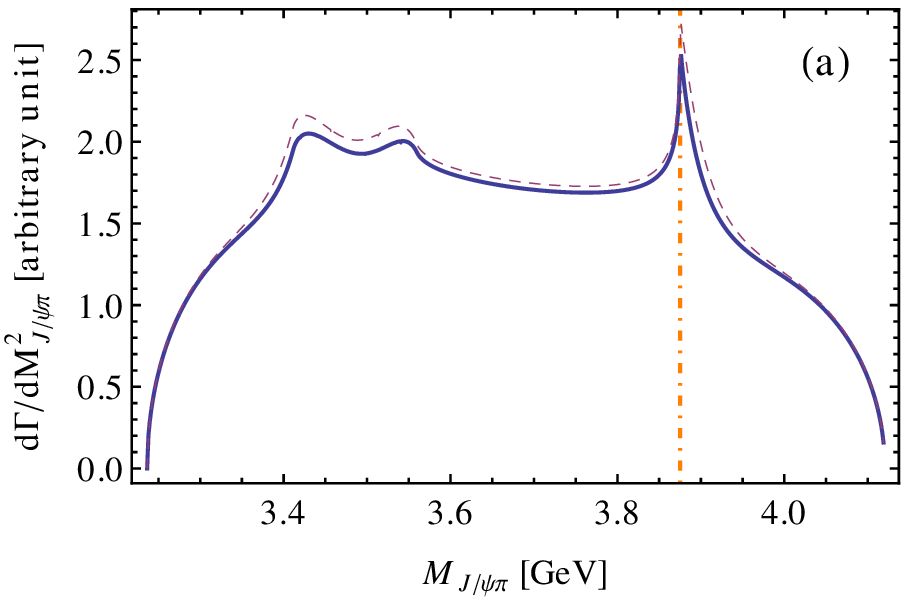}\end{minipage}\begin{minipage}{8cm}\includegraphics[width=1.0\hsize]{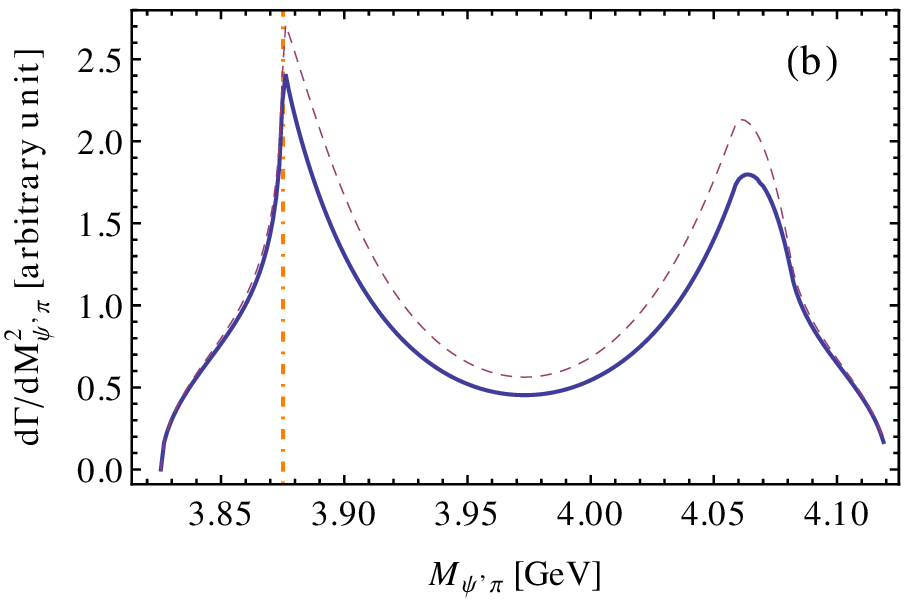}\end{minipage}
\caption{Invariant mass spectrum of $J/\psi\pi$($\psi^\prime\pi$)
in $Y(4260) \to D_1 \bar{D} [D^*]\to J/\psi(\psi^\prime) \pi\pi$.
Solid (dashed) line corresponds to the result with (without)
taking into account the width of $D_1$. The vertical dot-dashed
line indicates the the threshold of $\bar{D}D^*$.}\label{jpsipipi}
\end{figure}

\begin{figure}[tb]
\centering
\begin{minipage}{8cm}\includegraphics[width=1.0\hsize]{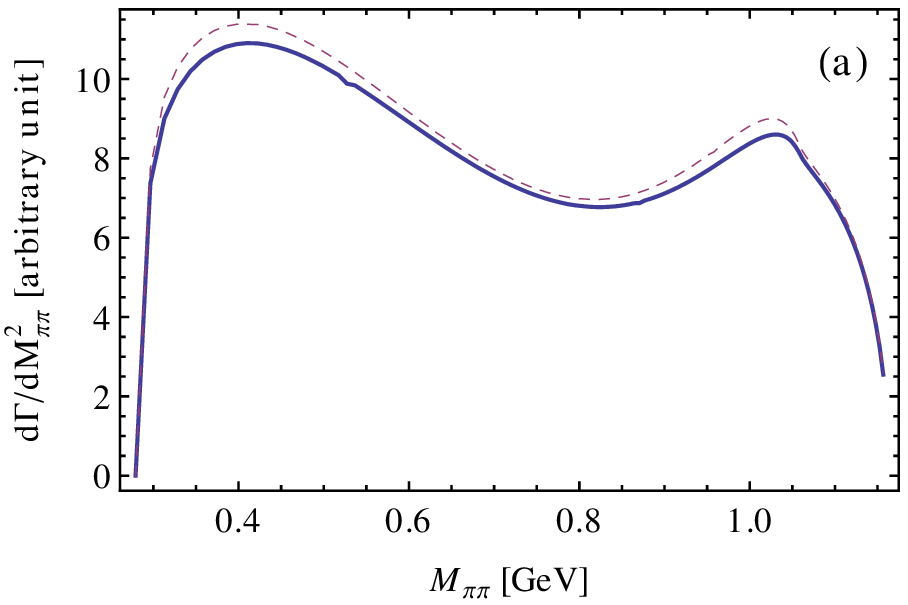}\end{minipage}\begin{minipage}{8cm}\includegraphics[width=1.0\hsize]{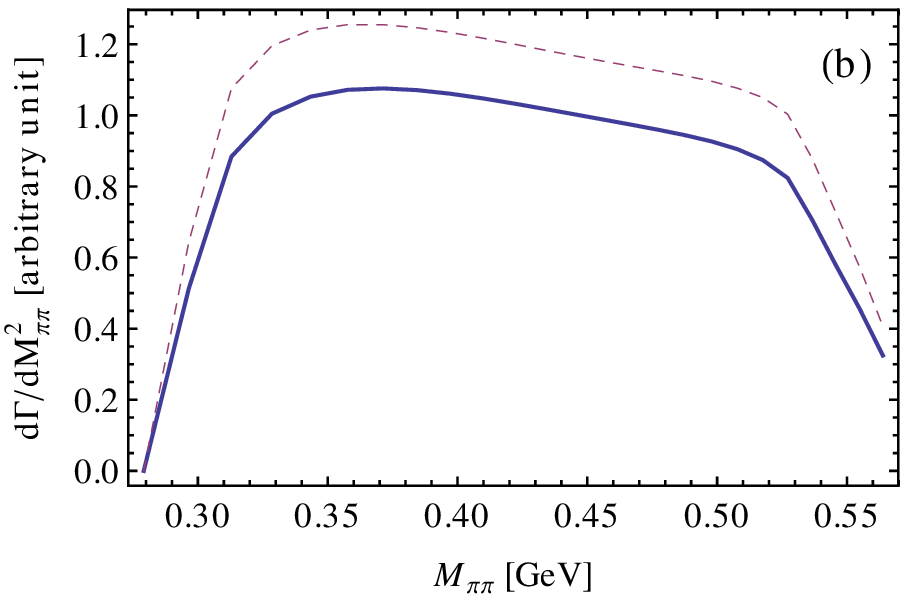}\end{minipage}
\caption{Invariant mass spectrum of $\pi\pi$ in (a): $Y(4260) \to
D_1 \bar{D} [D^*]\to J/\psi \pi\pi$, and (b): $Y(4260) \to D_1
\bar{D} [D^*]\to\psi^\prime \pi\pi$ . Solid (dashed) line
corresponds to the result with (without) taking into account the
width of $D_1$.}\label{IMSpipi}
\end{figure}

\begin{figure}[tb]
\centering
\begin{minipage}{8cm}\includegraphics[width=1.0\hsize]{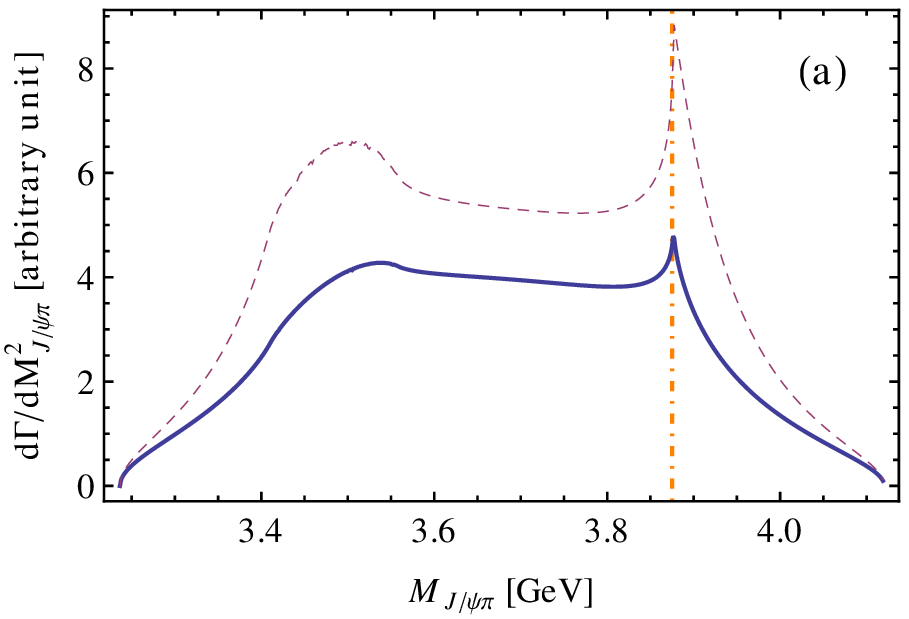}\end{minipage}\begin{minipage}{8cm}\includegraphics[width=1.0\hsize]{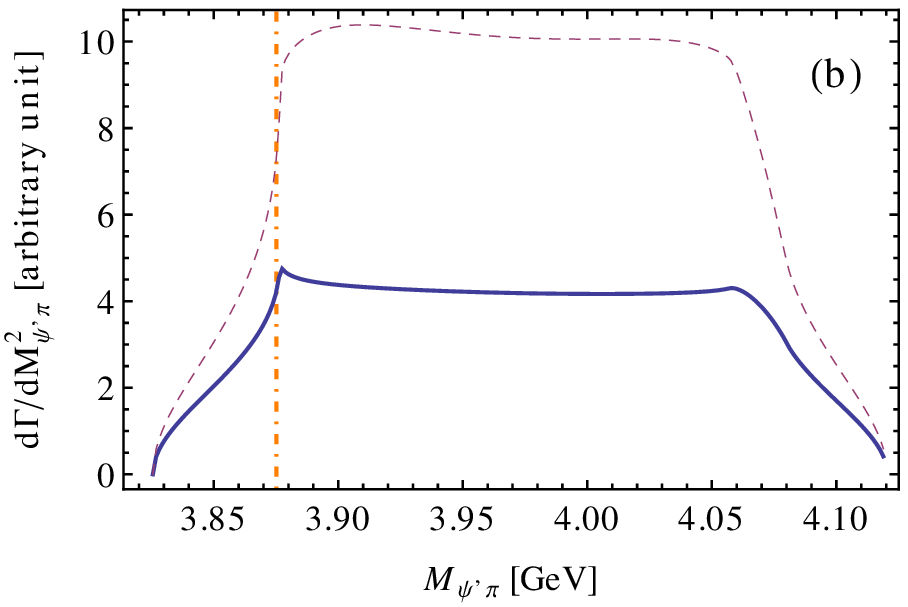}\end{minipage}
\caption{Invariant mass spectrum of $J/\psi\pi$($\psi^\prime\pi$)
in $Y(4260) \to D_0 \bar{D}^{*} [D]\to J/\psi(\psi^\prime)
\pi\pi$. Solid (dashed) line corresponds to the result with
(without) taking into account the width of $D_0$. The vertical
dot-dashed line indicates the the threshold of
$\bar{D}D^*$.}\label{D0jpsipi}
\end{figure}

\begin{figure}[tb]
\centering
\begin{minipage}{8cm}\includegraphics[width=1.0\hsize]{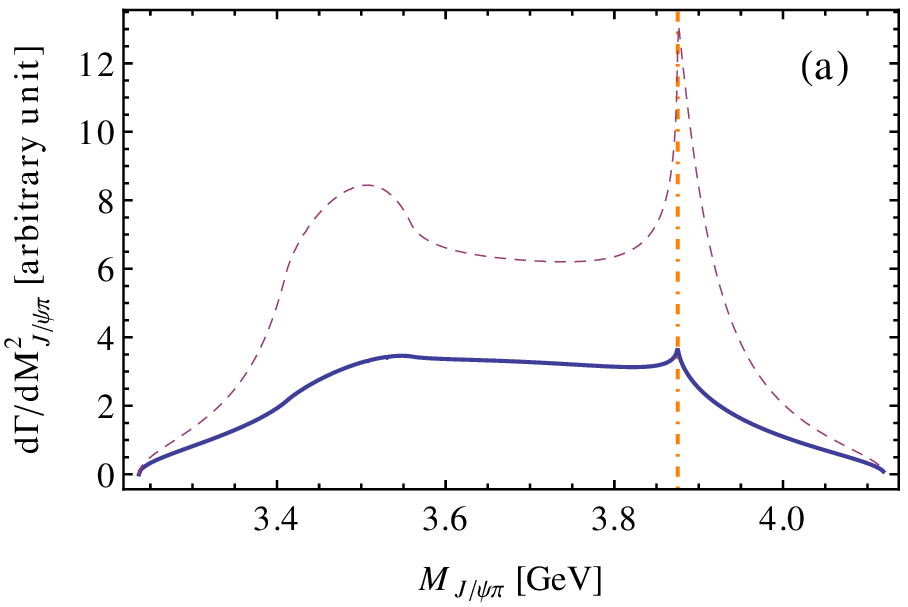}\end{minipage}\begin{minipage}{8cm}
\includegraphics[width=1.0\hsize]{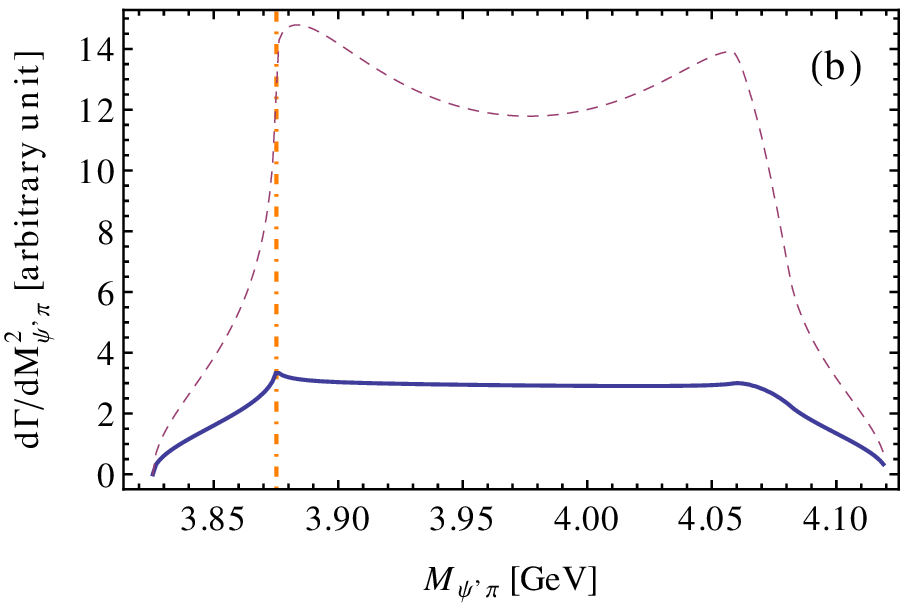}\end{minipage}
\caption{Invariant mass spectrum of $J/\psi\pi$($\psi^\prime\pi$)
in $Y(4260) \to D_1^{\prime} \bar{D} [D^*]\to J/\psi(\psi^\prime)
\pi\pi$. Solid (dashed) line corresponds to the result with
(without) taking into account the width of $D_1^\prime$. The
vertical dot-dashed line indicates the the threshold of
$\bar{D}D^*$.}\label{D1primejpsipi}
\end{figure}

As discussed in the previous section, the relative larger
$D^*\bar{D}$ yield in the vicinity of threshold will favor strong
FSI. In combination with the TS that may appear in the
rescattering amplitude, there will be a strong enhancement in
$J/\psi\pi$ ($\psi^\prime\pi$) invariant mass spectrum which lies
at the $D^*\bar{D}$ threshold. As illustrated in
Fig.~\ref{jpsipipi}, only according to this semi-quantitative
estimate, the lineshape behavior is already very similar with the
results observed by BESIII and BELLE. And the "width" of this cusp
structure can be compared with the width of "$Z_c(3900)$" to some
extent. While the $\pi\pi$ invariant mass spectrum would be
another story, as illustrated in Fig.~\ref{IMSpipi}. This rough
estimation can not give a very consistent distribution compared
with the data. Maybe after taking into account the $\pi\pi$ FSI,
this discrepancy could be compensated \cite{Dai:2012pb}.

Compared with $J/\psi\pi\pi$ final states, as the kinematics
changed, the reflection of the narrow cusp, which behaves as a
bump, has been shifted to the tail of the phase space.

The other possible combinations which may produce the similar cusp
structures are also investigated. Such as the processes  $Y(4260)
\to D_0 \bar{D}^{*} [D]\to J/\psi(\psi^\prime) \pi\pi$ and
$Y(4260) \to D_1^{\prime} \bar{D} [D^*]\to J/\psi(\psi^\prime)
\pi\pi$, where the charmed meson $D$ ($D^*$) in the bracket
denotes the exchanging particles between $D_0$ and $\bar{D}^*$
($D_1^\prime$ and $\bar{D}$). The numerical results are displayed
in Fig.~\ref{D0jpsipi} and Fig.~\ref{D1primejpsipi}. It seems that
there will also be obvious cusp structures in $J/\psi\pi\pi$ decay
mode if ignoring the larger width of intermediate states. As the
situation in $Y(4260) \to \bar{D}D^*\pi$, the broad width will
lower the amplitude, and the cusp structure would be smoothed out
to some extent. However if $Y(4260)$ has sizeable couplings with
$D_0 \bar{D}^*$ and $D_1^\prime \bar{D}$, these structures still
could offer some background to the pertinent decay modes.

The threshold of $\psi^\prime\pi$ is close to that of
$\bar{D}D^*$. If comparing Fig.~\ref{D0jpsipi}(b) or
Fig.~\ref{D1primejpsipi}(b) with Fig.~\ref{jpsipipi}(b), it seems
that only TS itself cannot give an obvious cusp, since these
distributions are results after integrating over phase space.
Generating an obvious cusp also require more $\bar{D}D^*$ events
should be produced at the threshold. The $D$-wave coupling of $D_1
D^*\pi$ or introducing a resonance $Z_c(3900)$, both of them can
make it work. From this point of view, it seems that the
$D_1\bar{D}$ combination is more favorable to be taken as the main
component of $Y(4260)$. The $\psi^\prime\pi\pi$ decay mode could
be used to test this argument.

\section{Threshold behavior in $Y(4260)$ radiative decays }
\begin{figure}[tb]
  \centering
  \includegraphics[width=0.8\hsize]{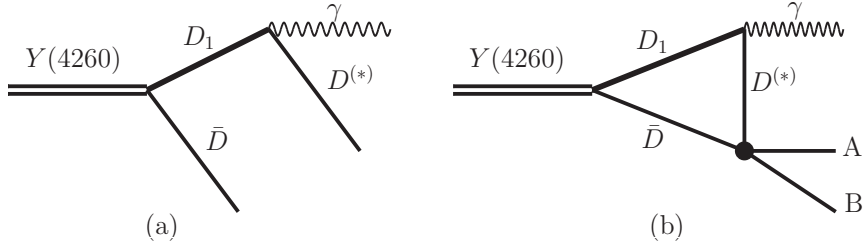}\\
  \caption{(a): Radiative decays of $Y(4260)$ by supposing it is a $D_1\bar{D}$ molecular state. (b): the rescattering process, where A and B are some specified final states.}\label{radiativediag}
\end{figure}

\begin{table}
\begin{center}
 \begin{tabular}{c|c|c|c}
  \hline
  \hline
  $\Gamma  \ \ [\mbox{KeV}]$   & Ref.~\cite{Maeda:2012nx} & Ref.~\cite{Close:2005se} & Ref.~\cite{Godfrey:2005ww} \\
  \hline
  $D_1^{0}\to \gamma D^0$ & 532 & 769 & 574  \\
  $D_1^{\pm}\to \gamma D^\pm$ & 8.8 & 20 & 58  \\
  $D_1^{0}\to \gamma D^{*0}$ & 136 & 273 & 85 \\
  $D_1^{0}\to \gamma D^{*\pm}$ & 5.3 & 44 & 8.6 \\
  \hline
\end{tabular}
\end{center}
 \caption{Radiative decay width of $D_1$ according to quark model estimation.}\label{tab:D1radiative}
\end{table}

\begin{figure}[tb]
\centering
\begin{minipage}{8cm}\includegraphics[width=1.0\hsize]{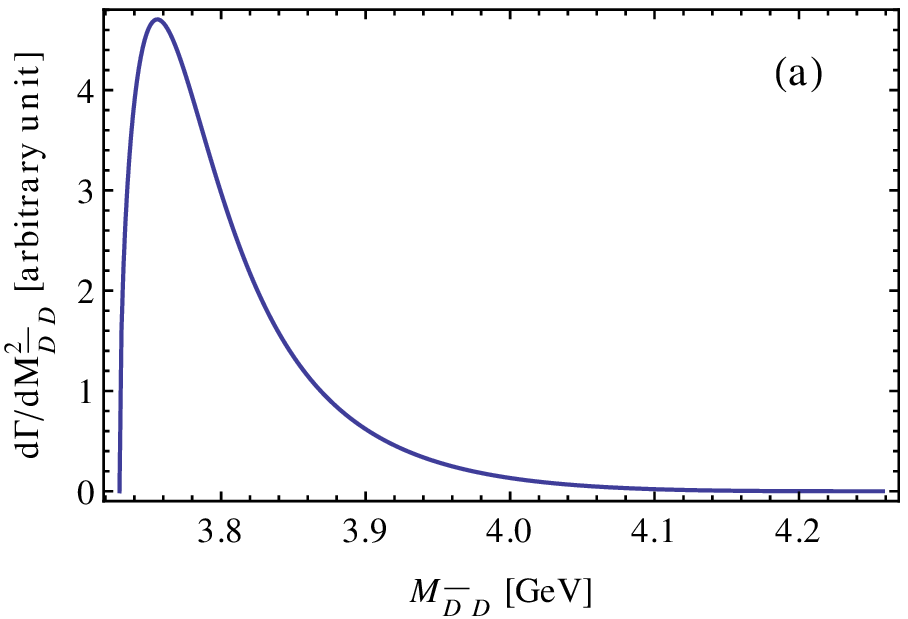}\end{minipage}\begin{minipage}{8cm}\includegraphics[width=1.0\hsize]{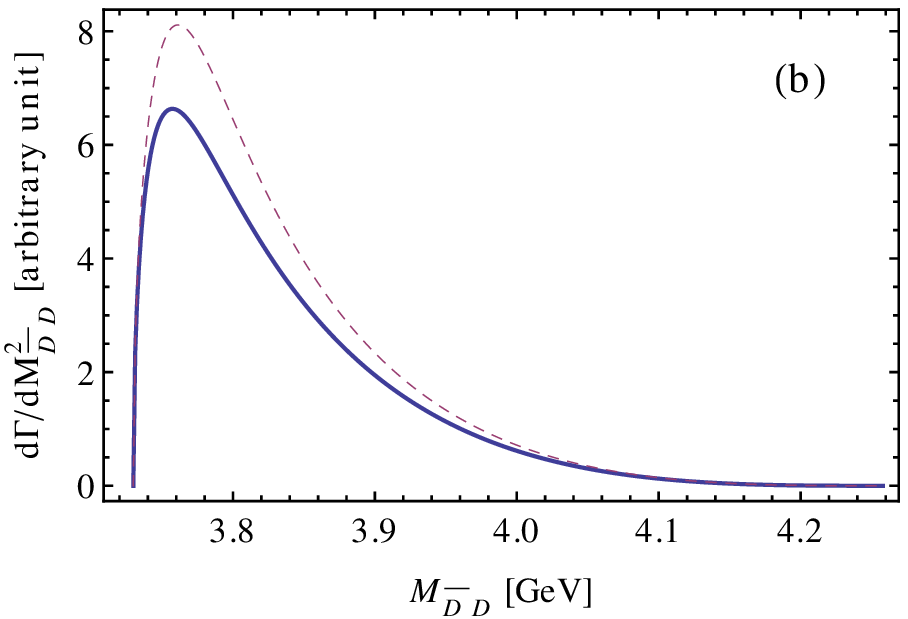}\end{minipage}
\caption{Invariant mass spectrum of $\bar{D}D$ in $Y(4260) \to D_1
\bar{D} \to \gamma D \bar{D}$. (a) and (b) correspond to the tree
and loop diagrams respectively. Solid (dashed) line corresponds to
the result with (without) taking into account the width of
$D_1$.}\label{radDD}
\end{figure}

\begin{figure}[tb]
\centering
\begin{minipage}{8cm}\includegraphics[width=1.0\hsize]{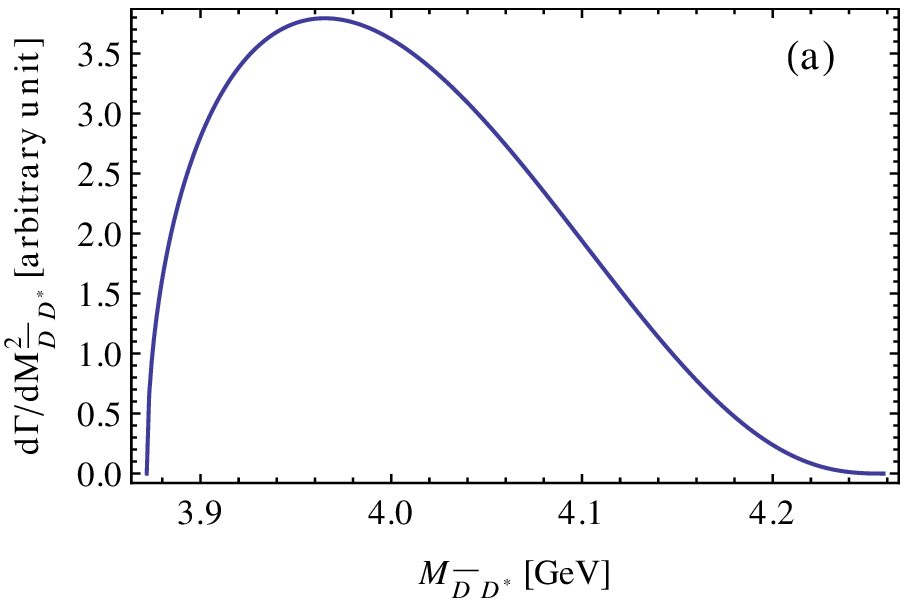}\end{minipage}\begin{minipage}{8cm}\includegraphics[width=1.0\hsize]{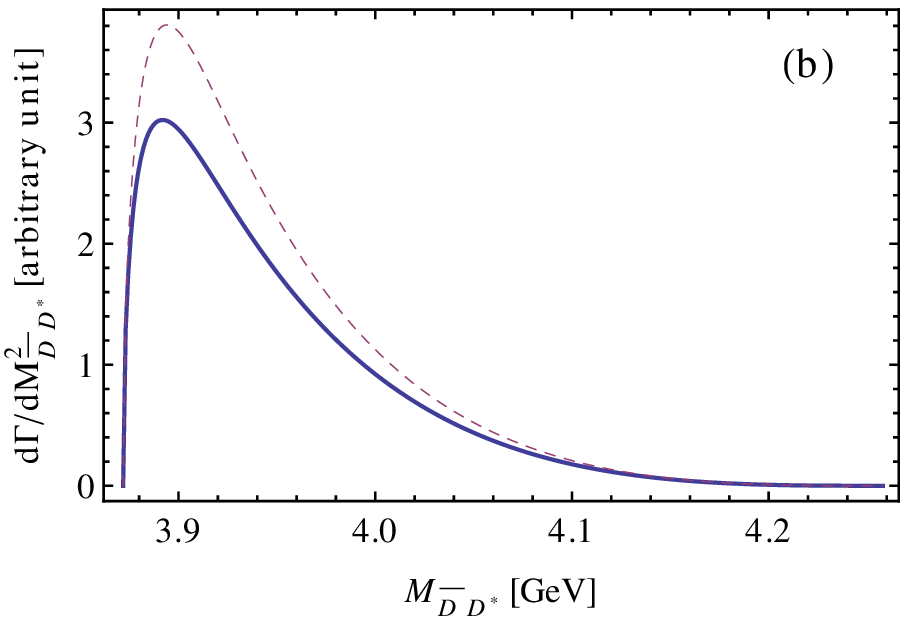}\end{minipage}
\caption{Invariant mass spectrum of $\bar{D}D^*$ in $Y(4260) \to
D_1 \bar{D} \to \gamma D^* \bar{D}$. (a) and (b) correspond to the
tree and loop diagrams respectively. Solid (dashed) line
corresponds to the result with (without) taking into account the
width of $D_1$.}\label{radDDstar}
\end{figure}

\begin{figure}[tb]
\centering
\begin{minipage}{8cm}\includegraphics[width=1.0\hsize]{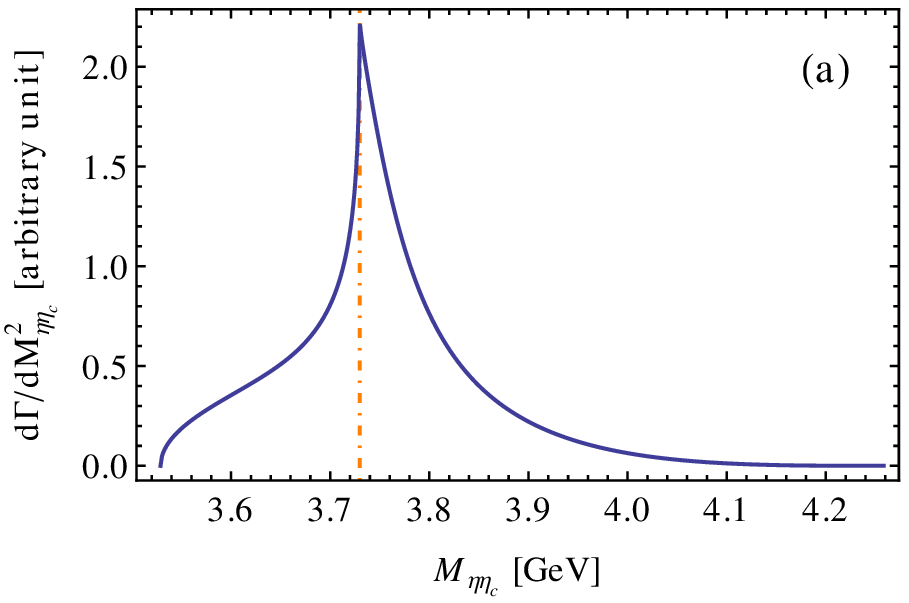}\end{minipage}\begin{minipage}{8cm}\includegraphics[width=1.0\hsize]{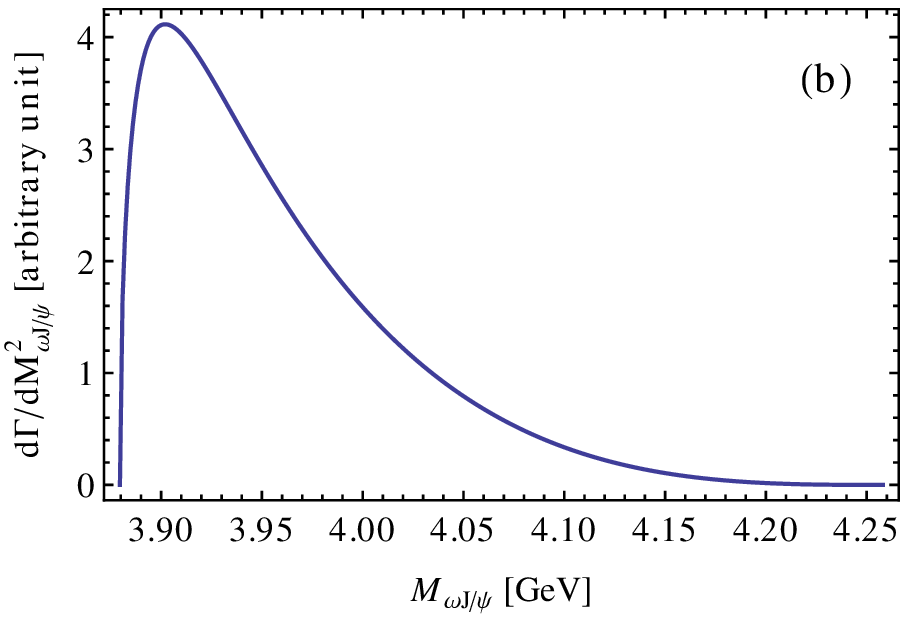}\end{minipage}
\caption{(a): Invariant mass spectrum of $\eta_c\eta$ in $Y(4260)
\to D_1 \bar{D} [D]\to \gamma \eta_c\eta$. The vertical dot-dashed
line corresponding to the threshold $2m_D$. (b): Invariant mass
spectrum of $J/\psi\omega$ in $Y(4260) \to D_1 \bar{D} [D^*]\to
\gamma J/\psi\omega$.}\label{etaceta}
\end{figure}

Radiative decay modes are very useful for understanding the
intrinsic structures of hadrons. By assuming $Y(4260)$ is a $D_1
\bar{D}$ molecule, we try to investigate some of its radiative
decay properties under this scenario. As illustrated in
Fig.~\ref{radiativediag}, we will also concentrate on the
threshold phenomena resulted from the rescattering process. Since
the kinematics is similar with the situation discussed in the
previous sections, TS is also expected to occur, which may
strongly enhance the amplitude and change the corresponding
lineshape behavior.  The Lagrangian describing $D_1$ radiative
decays reads
\begin{eqnarray}
\mathcal{L}_{EM}=i\bar{\mu} \ Tr[\bar{H}_a T^\mu_b \gamma^\nu
F_{\mu\nu}]Q_{ba},
\end{eqnarray}
where $Q=\mbox{diag}(2/3,-1/3,-1/3)$, $F_{\mu\nu}=\partial_\mu
A_\nu-\partial_\nu A_\mu$, and $\bar{\mu}$ is the coupling
strength. It should be mentioned the spin symmetry violating
operators and the contact interactions are not contained in this
Lagrangian. Since for the radiative decay modes discussed here, we
will mainly pay attention to the lineshape behavior close to the
threshold. Only the diagrams where TS may occur in the transition
amplitude are considered, while some other diagrams which can be
taken as background are ignored. According to the results
estimated by utilizing quark model, which are displayed in
Tab.~\ref{tab:D1radiative}, the radiative decay width of
$D_1^0\to\gamma D^{(*)0}$ is much larger than that of
$D_1^\pm\to\gamma D^{(*)\pm}$. This may imply that the ratio $R
\equiv \Gamma(Y(4260)\to\gamma D^{(*)0}\bar{D}^0) /
\Gamma(Y(4260)\to\gamma D^{(*)+} D^-)$ will be much larger than 1.
And the rescattering will mainly happen between neutral charmed
anti-charmed meson pairs. However, if FSI are strong, or there is
a larger branching ratio of $Y\to \gamma X\to \gamma
D^{(*)}\bar{D}$, the ratio will be changed, where $X$ is some kind
of charmonium state, such as $\chi_{cJ}$ or $X(3872)$.

With the similar discussion as the previous sections, taking the
final states $AB$ as $D\bar{D}$ and $D^*\bar{D}$ respectively, the
numerical results without taking into account interference are
displayed in Fig.~\ref{radDD} and Fig.~\ref{radDDstar}. It can be
noticed that in both of these two decay modes, the rescattering
process can give obvious threshold enhancement. If taking $AB$ as
$\eta\eta_c$ and $\omega J/\psi$ respectively, and assuming an
$S$-wave coupling, a narrow cusp structure is obtained at
$D\bar{D}$ as illustrated in  Fig.~\ref{etaceta}(a). Since the
threshold of $\omega J/\psi$ is a little higher than that of
${D}^{*} \bar{D}$, there is only a threshold enhancement at
Fig.~\ref{etaceta}(b). These resonance like structures do not
result from some genuine resonances with quantum numbers $J^P=0^+$
and $J^P=1^+$, but due to TS could occur in the rescattering
amplitude, the narrow cusp structure is also expected. Especially
for the narrow cusp structure in the vicinity of $D\bar{D}$
threshold, since there is no scalar charmonium state with the mass
about $2m_D$ that has ever been observed, this radiative decay
mode can be taken as a criterion to test whether such a structure
is some kind of molecular state or just a cusp structure results
from FSI.

\section{Summary}
In this paper by assuming $Y(4260)$ is a $D_1\bar{D}$ molecular
state, we investigate several decay modes which are related with
this assumption. $\bar{D}D^*\pi$ will be the main decay channel
under this ansatz, and since $D_1\to D^*\pi$ is $D$-wave decays,
there will be more $\bar{D}D^*$ events accumulated in the vicinity
of threshold. Therefore, strong FSI would be expected. With some
special kinematic configurations, TS may occur in the transition
amplitude, which will significantly change the threshold behavior
and manifest itself as some cusp structures. The process
$Y(4260)\to J/\psi(\psi^\prime) \pi\pi$ results from $\bar{D}D^*$
rescattering has been investigated, where the cusp structure is
very similar with the experimental observations. This is the
result without introducing an apparent true resonance, such as
$Z_c(3900)$. However, we should also claim this effect just offer
a  possible dynamical mechanism to describe such a resonance like
structure, but not exclude the existence of a genuine resonance.
And the other possible combinations $D_0 \bar{D}^*$ and
$D_1^\prime \bar{D}$, which are close to the mass of $Y(4260)$,
have also been taken into account. Although they are too broad to
form a molecular state, they can also lead to some threshold
enhancement structures, as long as they have sizeable couplings
with the initial state. Some radiative decay modes of $Y(4260)$
under the molecule ansatz have also been discussed. And we
emphasize that the strong $D\bar{D}$ $S$-wave interaction will
lead to a narrow cusp structure in the process $Y(4260) \to D_1
\bar{D} [D]\to \gamma \eta_c\eta$, which may behave itself as a
scalar charmonium resonance.

Although the $D_1\bar{D}$ molecule description of $Y(4260)$ is
natural for explaining some of the experimental observations, this
ansatz is not necessary for the TS mechanism discussed in this
paper. As long as the state has a larger coupling with
$D_1\bar{D}$ and its mass is very close to the threshold, we can
expect the similar enhancement phenomena, no matter whatever it
is.

\subsection*{Acknowledgments}
We would like to thank S. L. Zhu, Q. Zhao and Z. W. Liu for useful
discussions. This work was supported in part by China Postdoctoral
Science Foundation (2013M530461), the National
Natural Science Foundation of China under Grants 11275113,
11075004, 11021092 and 11261130311, and the DFG and the NSFC
through funds provided to the sinogermen CRC 110 "Symmetries
and the Emergence of Structure in QCD".

.

\end{document}